\documentclass[fleqn,twocolumn,showpacs,preprintnumbers,showkeys]{revtex4}
\usepackage{graphicx}
\usepackage{amssymb}
\usepackage{amsmath}
\usepackage{bm}
\usepackage[flushleft]{caption2}

\begin{document}
\setcounter{page}{1}
\title{Forces in electromagnetic field and gravitational field}
\author{Zihua Weng}
\email{xmuwzh@xmu.edu.cn.}
\affiliation{School of Physics and
Mechanical \& Electrical Engineering,
\\Xiamen University, Xiamen 361005, China}

\begin{abstract}
The force can be defined from the linear momentum in the
gravitational field and electromagnetic field. But this definition
can not cover the gradient of energy. In the paper, the force will
be defined from the energy and torque in a new way, which involves
the gravitational force, electromagnetic force, inertial force,
gradient of energy, and some other new force terms etc. One of these
new force terms can be used to explain why the solar wind varies
velocity along the magnetic force line in the interplanetary space
between the sun and the earth.
\end{abstract}



\maketitle

\section{INTRODUCTION}

The force is one important physical conception for the
electromagnetic and gravitational fields. There is only the gravity
\cite{newton}, Lorentz force \cite{maxwell}, and gradient of energy
in classical field theory described by the vector. But these forces
can not explain why the solar wind varies velocity along the
magnetic force line in the interplanetary space and the coronal hole
on the sun \cite{tu}.

Since the antiquity, scientists have used the concept of force in
the study of different physical objects. Modern description of force
was introduced by I. Newton in the 17th century. Nowadays, the
description of force covers various forces in the gravitational and
electromagnetic fields \cite{morita}. In the paper, making use of
quaternions \cite{hamilton, adler}, all kinds of forces we found up
to now can be described with one single formula, including the
gravity, Lorentz force, and gradient of energy etc. Meanwhile all
kinds of energies we knew at present can also be described with one
single formula, including the work, kinetic energy, potential energy,
field energy, proper energy, interacting energy among magnetic
fields with magnetic moments, and some other new energy terms etc.

In the electromagnetic field and gravitational field, the force can
be defined from the linear momentum \cite{weng}. The forces include
the inertial force, gravitational force, and Lorentz force, except
for the gradient of energy. Further, the force can be defined from
the energy and torque in a new way. The force is able to involve the
gravity, inertial force, Lorentz force, gradient of energy, and some
other new force terms etc. And one of these new force terms can explain
the dynamical puzzle of solar wind.

\section{Forces from linear momenta}

The forces can be defined from the linear momentum in the
electromagnetic field and gravitational field. And the forces
include the inertial force, gravitational force, electric force, and
magnetic force, etc.

\subsection{Gravitational field}

The force in the gravitational field can be described by
quaternions. In the quaternion space, the coordinates are $r_0$,
$r_1$, $r_2$, and $r_3$, with the basis vector $\mathbb{E}_g = (1 ,
\emph{\textbf{i}}_1 , \emph{\textbf{i}}_2 , \emph{\textbf{i}}_3)$.
Herein $r_0 = v_0 t$, $t$ is the time; $v_0$ is the speed of
gravitational intermediate boson, which is the first part of the
photon. The radius vector is $\mathbb{R}_g = r_0 + \Sigma (r_j
\emph{\textbf{i}}_j)$, and the velocity is $\mathbb{V}_g = v_0 +
\Sigma (v_j \emph{\textbf{i}}_j)$, $j = 1, 2, 3$; $i = 0, 1, 2, 3 $.

\subsubsection{Linear momentum in gravitational field}

In the quaternion space for the gravitational field, the gravitational potential $\mathbb{A}_g$ is
\begin{eqnarray}
\mathbb{A}_g = a_0 + \Sigma (a_j \emph{\textbf{i}}_j)~,
\end{eqnarray}
and the strength $\mathbb{B}_g$ of gravitational field
\begin{eqnarray}
\mathbb{B}_g = \lozenge \circ \mathbb{A}_g = b_0 + \Sigma (b_j \emph{\textbf{i}}_j)~,
\end{eqnarray}
where $a_0$ is the scalar potential, while $\textbf{a}$ is the vectorial potential.
The $\circ$ denotes the quaternion multiplication. $ \lozenge
= \partial_0 + \Sigma (\emph{\textbf{i}}_j
\partial_j) ; ~\partial_i = \partial/\partial r_i$;
$\textbf{a} = \Sigma (a_j \emph{\textbf{i}}_j)$; $\nabla = \Sigma
(\emph{\textbf{i}}_j \partial_j)$. The gauge is selected as $b_0 = \partial_0 a_0 + \nabla \cdot
\textbf{a} = 0$ .

\begin{table}[b]
\caption{\label{tab:table1}The quaternion multiplication table.}
\begin{ruledtabular}
\begin{tabular}{ccccc}
$ $ & $1$ & $\emph{\textbf{i}}_1$  & $\emph{\textbf{i}}_2$ &
$\emph{\textbf{i}}_3$  \\
\hline $1$ & $1$ & $\emph{\textbf{i}}_1$  & $\emph{\textbf{i}}_2$ &
$\emph{\textbf{i}}_3$  \\
$\emph{\textbf{i}}_1$ & $\emph{\textbf{i}}_1$ & $-1$ &
$\emph{\textbf{i}}_3$  & $-\emph{\textbf{i}}_2$ \\
$\emph{\textbf{i}}_2$ & $\emph{\textbf{i}}_2$ &
$-\emph{\textbf{i}}_3$ & $-1$ & $\emph{\textbf{i}}_1$ \\
$\emph{\textbf{i}}_3$ & $\emph{\textbf{i}}_3$ &
$\emph{\textbf{i}}_2$ & $-\emph{\textbf{i}}_1$ & $-1$
\end{tabular}
\end{ruledtabular}
\end{table}

The gravitational strength $\mathbb{B}_g$ includes two kinds of
components $\textbf{g}/v_0 = \partial_0 \textbf{a} + \nabla a_0 $
and $\textbf{b} = \nabla \times \textbf{a} $ ,
\begin{eqnarray}
\textbf{g}/v_0 = && \emph{\textbf{i}}_1 ( \partial_0 a_1 +
\partial_1 a_0 ) + \emph{\textbf{i}}_2 ( \partial_0 a_2 + \partial_2
a_0 )
\nonumber\\
&& + \emph{\textbf{i}}_3 ( \partial_0 a_3 + \partial_3 a_0 )~,
\\
\textbf{b} = && \emph{\textbf{i}}_1 ( \partial_2 a_3 -
\partial_3 a_2 ) + \emph{\textbf{i}}_2 ( \partial_3 a_1 - \partial_1
a_3 )
\nonumber\\
&& + \emph{\textbf{i}}_3 ( \partial_1 a_2 - \partial_2 a_1 )~,
\end{eqnarray}
where the $\textbf{b}$ may be too weak to be detected presently; and there are $\textbf{a} = 0$ and $\textbf{b} = 0$ in Newtonian gravitational theory. While the $\textbf{g}$ and $\textbf{b}$ are corresponded to the linear acceleration and angular velocity respectively.

The gravitational source $\mathbb{S}$ includes the linear
momentum density $\mathbb{S}_g = m \mathbb{V}_g $ and an extra part
$v_0 \triangle m_1$ ,
\begin{eqnarray}
\mu \mathbb{S} = - ( \mathbb{B}_g/v_0 + \lozenge)^* \circ \mathbb{B}_g
= \mu_g (\mathbb{S}_g + v_0 \triangle m_1 )~,
\end{eqnarray}
where $m$ is the mass density; $*$ denotes the quaternion
conjugate; $\mu$ is one coefficient, and $\mu_g$ is the
gravitational constant; $\mathbb{B}_g^* \circ \mathbb{B}_g/(2\mu_g)$
is the energy density of gravitational field; $\triangle m_1 = -
\mathbb{B}_g^* \circ \mathbb{B}_g/(\mu_g v_0^2)$ .

The force density $\mathbb{F}_g$ is defined from the linear momentum
density $\mathbb{P}_g = \mu \mathbb{S} / \mu_g$ . And the latter is
the extension of the $\mathbb{S}_g$ .
\begin{eqnarray}
\mathbb{F}_g = v_0 (\mathbb{B}_g / v_0 + \lozenge )^* \circ
\mathbb{P}_g~.
\end{eqnarray}

We introduce above definition so as to recover various forces
regarding gravitational fields including the inertial force
density and gravity density etc.

\begin{table}[t]
\caption{\label{tab:table1}The operator and multiplication of the
physical quantity in the quaternion space.}
\begin{ruledtabular}
\begin{tabular}{ll}
$definitions$               &   $meanings$                                            \\
\hline
$\nabla \cdot \textbf{a}$   &  $-(\partial_1 a_1 + \partial_2 a_2 + \partial_3 a_3)$  \\
$\nabla \times \textbf{a}$  &  $\emph{\textbf{i}}_1 ( \partial_2 a_3
                                 - \partial_3 a_2 ) + \emph{\textbf{i}}_2 ( \partial_3 a_1
                                 - \partial_1 a_3 )$                                  \\
$$                          &  $ + \emph{\textbf{i}}_3 ( \partial_1 a_2
                                 - \partial_2 a_1 )$                                  \\
$\nabla a_0$                &  $\emph{\textbf{i}}_1 \partial_1 a_0
                                 + \emph{\textbf{i}}_2 \partial_2 a_0
                                 + \emph{\textbf{i}}_3 \partial_3 a_0  $              \\
$\partial_0 \textbf{a}$     &  $\emph{\textbf{i}}_1 \partial_0 a_1
                                 + \emph{\textbf{i}}_2 \partial_0 a_2
                                 + \emph{\textbf{i}}_3 \partial_0 a_3  $              \\
\end{tabular}
\end{ruledtabular}
\end{table}

\begin{table}[b]
\caption{\label{tab:table1}Some definitions and the gravitational
force density in the quaternion space.}
\begin{ruledtabular}
\begin{tabular}{ll}
$ definitions $                                 & $ meanings $ \\
\hline
$v_0 \partial_0 \textbf{p}$                     & inertial force density\\
$m \textbf{g}^*$                                & gravity density  \\
$\triangle m_1 \textbf{g}^*$                      & new force part \\
$p_0 \textbf{b}^* $                             & new force part \\
$v_0 \nabla^* p_0$                              & new force part \\
$\textbf{g}^* \times \textbf{p}/v_0$            & new force part \\
$\textbf{b}^* \times \textbf{p}$                & new force part \\
$v_0 \nabla^* \times \textbf{p}$                & new force part \\
\end{tabular}
\end{ruledtabular}
\end{table}

\subsubsection{Gravitational force}

In the quaternion space, the inertial mass density is $m$, and the
gravitational mass density is $\widehat{m} = m + \triangle m_1$. The
linear momentum density is $\mathbb{P}_g = p_0 + \Sigma (p_j
\emph{\textbf{i}}_j )$. Herein $p_0 = \widehat{m} v_0$ , $p_j = m
v_j $ , and $\textbf{p} = \Sigma (p_j \emph{\textbf{i}}_j) $ .

By Eq.(6), the gravitational force density $\mathbb{F}_g$ is
\begin{eqnarray}
\mathbb{F}_g = f_0 + \Sigma (f_j \emph{\textbf{i}}_j )~,
\end{eqnarray}
where $f_0 = \partial p_0 / \partial t + v_0 \Sigma ( \partial p_j
/ \partial r_j ) + \Sigma ( b_j p_j ) $ .

In the quaternion space, the vectorial part $\textbf{f} = \Sigma
(f_j \emph{\textbf{i}}_j )$ of force density $\mathbb{F}_g$ can be
decomposed from Eq.(7) ,
\begin{eqnarray}
\textbf{f} = && v_0 \partial_0 \textbf{p} + p_0 \textbf{g}^*/v_0 +
p_0 \textbf{b}^*
\nonumber\\
&& + v_0 \nabla^* p_0 + (\textbf{g}/v_0 + \textbf{b} + v_0 \nabla)^*
\times \textbf{p}~,
\end{eqnarray}
where $ p_0 \textbf{g}^*/v_0 = m \textbf{g}^* + \triangle m_1 \textbf{g}^* $, and $m \textbf{g}^*$ is the
gravity density; $[\triangle m_1 \textbf{g}^* + v_0 \nabla^* p_0 + p_0 \textbf{b}^* + (\textbf{g}/v_0 + \textbf{b} + v_0 \nabla)^* \times
\textbf{p}]$ is one new part of gravitational force density; $v_0 \partial_0 \textbf{p}$ is the inertial force density.

In the gravitational field, some new parts of above force density
are too weak to be detected. As a particular case, these new parts
of forces are equal to zero in Newtonian gravitational theory.

\subsection{Electromagnetic field}

The electric and magnetic forces can be described by the algebra of
quaternions. In the quaternion space, the coordinates are
$r_0$, $r_1$, $r_2$, and $r_3$, with the basis vector $\mathbb{E}_q
= (1 , \emph{\textbf{i}}_1 , \emph{\textbf{i}}_2 ,
\emph{\textbf{i}}_3)$. Herein $r_0 = v_0 t$,  $t$ is the time; $v_0$
is the speed of electromagnetic intermediate boson, which is the
second part of the photon. The radius vector is $\mathbb{R}_q = r_0
+ \Sigma (r_j \emph{\textbf{i}}_j)$, and the velocity is
$\mathbb{V}_q = v_0 + \Sigma (v_j \emph{\textbf{i}}_j)$.

\subsubsection{Linear momentum in electromagnetic field}

In the quaternion space for the electromagnetic field, the electromagnetic potential $\mathbb{A}_q$ is
\begin{eqnarray}
\mathbb{A}_q = A_0 + \Sigma (A_j \emph{\textbf{i}}_j)~,
\end{eqnarray}
and the strength $\mathbb{B}_q$ of electromagnetic field
\begin{eqnarray}
\mathbb{B}_q = \lozenge \circ \mathbb{A}_q = B_0 + \Sigma (B_j \emph{\textbf{i}}_j)~.
\end{eqnarray}

The electromagnetic strength $\mathbb{B}_q$ includes two parts,
$\textbf{E}/v_0 = \partial_0 \textbf{A} + \nabla A_0 $ and
$\textbf{B} = \nabla \times \textbf{A} $ ,
\begin{eqnarray}
\textbf{E}/v_0 = && \emph{\textbf{i}}_1 ( \partial_0 A_1 +
\partial_1 A_0 ) + \emph{\textbf{i}}_2 ( \partial_0 A_2 + \partial_2
A_0 )
\nonumber\\
&& + \emph{\textbf{i}}_3 ( \partial_0 A_3 + \partial_3 A_0 )~,
\\
\textbf{B} = && \emph{\textbf{i}}_1 ( \partial_2 A_3 -
\partial_3 A_2 ) + \emph{\textbf{i}}_2 ( \partial_3 A_1 - \partial_1
A_3 )
\nonumber\\
&& + \emph{\textbf{i}}_3 ( \partial_1 A_2 - \partial_2 A_1 )~,
\end{eqnarray}
where the gauge is selected as $B_0 = \partial_0 A_0 + \nabla \cdot \textbf{A} = 0$, with $\textbf{A} = \Sigma (A_j \emph{\textbf{i}}_j)$. Similarly $\textbf{E}$ and $\textbf{B}$ are corresponded to the linear-like and angular-like quantity respectively. And the latter may be associated with the spin.

The electromagnetic source $\mathbb{S}$ encompasses the
electric current density $\mathbb{S}_q = q \mathbb{V}_q $ and an
extra part $v_0 \triangle m_2 $,
\begin{eqnarray}
\mu \mathbb{S} = && - (k_{eg} \mathbb{B}_q/v_0 + \lozenge)^* \circ
( k_{eg} \mathbb{B}_q )
\nonumber \\
= && k_{eg} \mu_q \mathbb{S}_q - k_{eg}^2 \mathbb{B}_q^* \circ
\mathbb{B}_q/v_0~,
\end{eqnarray}
where $q$ is the charge density; $\mu_q$ is the electromagnetic
constant; $k_{eg}$ is a coefficient; $\mathbb{B}_q^* \circ \mathbb{B}_q/(2\mu_q)$ is the electromagnetic energy
density; $\triangle m_2 = - \mathbb{B}_q^* \circ \mathbb{B}_q/(\mu_q v_0^2)$ .

The force density $\mathbb{F}_q$ is defined from the linear momentum
density $\mathbb{P}_q = \mu \mathbb{S} / \mu_g$ , that is
\begin{eqnarray}
\mathbb{F}_q = v_0 (k_{eg} \mathbb{B}_q/v_0 + \lozenge )^* \circ
\mathbb{P}_q~,
\end{eqnarray}
where the force density includes the inertial force density and
electromagnetic force density, etc.

\begin{table}[t]
\caption{\label{tab:table1}Some definitions and the electromagnetic
force density in the quaternion space.}
\begin{ruledtabular}
\begin{tabular}{ll}
$ definitions $                                 & $ meanings $ \\
\hline
$v_0 \partial_0 \textbf{P}$                     & (inertial force density)\\
$q \textbf{E}^*$                                & electric force density\\
$q \textbf{B}^* \times \textbf{V}$              & magnetic force density\\
$k_{eg} \triangle m_2 \textbf{E}^*/v_0$           & (new force part) \\
$q v_0 \textbf{B}^* $                           & new force part \\
$k_{eg} \triangle m_2 \textbf{B}^* $              & (new force part) \\
$v_0 \nabla^* P_0$                              & (new force part) \\
$q \textbf{E}^* \times \textbf{V}/v_0$          & new force part \\
$v_0 \nabla^* \times \textbf{P}$                & (new force part) \\
\end{tabular}
\end{ruledtabular}
\end{table}

\subsubsection{Electromagnetic force}

In the quaternion space, the linear momentum density is
$\mathbb{P}_q = P_0 + \Sigma (P_j \emph{\textbf{i}}_j )$, with $P_i
= M V_i $ , $M = k_{eg} q \mu_q / \mu_g$ .

By Eq.(6), the electromagnetic force density $\mathbb{F}_q$ is
\begin{eqnarray}
\mathbb{F}_q = F_0 + \Sigma (F_j \emph{\textbf{i}}_j )~,
\end{eqnarray}
where $F_0 = \partial P_0 / \partial t + v_0 \Sigma (  \partial P_j
/ \partial r_j ) + \Sigma ( k_{eg} B_j P_j ) $ .

The vectorial part $\textbf{f}_q = \Sigma (F_j \emph{\textbf{i}}_j)$
of force density $\mathbb{F}_q$ can be decomposed from Eq.(15).
\begin{eqnarray}
\textbf{f}_q = && v_0 \partial_0 \textbf{P} + k_{eg} P_0
\textbf{E}^*/v_0 + k_{eg} P_0 \textbf{B}^* + v_0 \nabla^* P_0
\nonumber\\
&& + ( k_{eg} \textbf{E}/v_0 + k_{eg} \textbf{B} + v_0 \nabla)^*
\times \textbf{P}~,
\end{eqnarray}
where $\textbf{P} = \Sigma (P_j \emph{\textbf{i}}_j) $; $\textbf{V}
= \Sigma (V_j \emph{\textbf{i}}_j) $; $v_0 \partial_0 \textbf{P}$ is
the inertial force density; $k_{eg} P_0 \textbf{E}^* /v_0 = q
\textbf{E}^* + k_{eg} \triangle m_2 \textbf{E}^*/v_0 $; $q
\textbf{E}^*$ is the electric force density; $ k_{eg} \textbf{B}^*
\times \textbf{P} = q \textbf{B}^* \times \textbf{V}$ is the
magnetic force density; $[ k_{eg} \triangle m_2 \textbf{E}^* / v_0 +
v_0 \nabla^* P_0 + k_{eg} P_0 \textbf{B}^* + ( k_{eg} \textbf{E}/v_0
+ v_0 \nabla)^* \times \textbf{P}]$ is one new part of the
electromagnetic force density.

The $q v_0 \textbf{B}^* $ is one new force
term, which can be used to explain why the solar wind varies the
velocity along the magnetic force line. Other new force terms may be too weak to be detected
in the weak electromagnetic field. Moreover some terms with the parentheses should not be physical in Table IV.

\begin{table}[b]
\caption{\label{tab:table1}The octonion multiplication table.}
\begin{ruledtabular}
\begin{tabular}{ccccccccc}
$ $ & $1$ & $\emph{\textbf{i}}_1$  & $\emph{\textbf{i}}_2$ &
$\emph{\textbf{i}}_3$  & $\emph{\textbf{I}}_0$  &
$\emph{\textbf{I}}_1$
& $\emph{\textbf{I}}_2$  & $\emph{\textbf{I}}_3$  \\
\hline $1$ & $1$ & $\emph{\textbf{i}}_1$  & $\emph{\textbf{i}}_2$ &
$\emph{\textbf{i}}_3$  & $\emph{\textbf{I}}_0$  &
$\emph{\textbf{I}}_1$
& $\emph{\textbf{I}}_2$  & $\emph{\textbf{I}}_3$  \\
$\emph{\textbf{i}}_1$ & $\emph{\textbf{i}}_1$ & $-1$ &
$\emph{\textbf{i}}_3$  & $-\emph{\textbf{i}}_2$ &
$\emph{\textbf{I}}_1$
& $-\emph{\textbf{I}}_0$ & $-\emph{\textbf{I}}_3$ & $\emph{\textbf{I}}_2$  \\
$\emph{\textbf{i}}_2$ & $\emph{\textbf{i}}_2$ &
$-\emph{\textbf{i}}_3$ & $-1$ & $\emph{\textbf{i}}_1$  &
$\emph{\textbf{I}}_2$  & $\emph{\textbf{I}}_3$
& $-\emph{\textbf{I}}_0$ & $-\emph{\textbf{I}}_1$ \\
$\emph{\textbf{i}}_3$ & $\emph{\textbf{i}}_3$ &
$\emph{\textbf{i}}_2$ & $-\emph{\textbf{i}}_1$ & $-1$ &
$\emph{\textbf{I}}_3$  & $-\emph{\textbf{I}}_2$
& $\emph{\textbf{I}}_1$  & $-\emph{\textbf{I}}_0$ \\
\hline $\emph{\textbf{I}}_0$ & $\emph{\textbf{I}}_0$ &
$-\emph{\textbf{I}}_1$ & $-\emph{\textbf{I}}_2$ &
$-\emph{\textbf{I}}_3$ & $-1$ & $\emph{\textbf{i}}_1$
& $\emph{\textbf{i}}_2$  & $\emph{\textbf{i}}_3$  \\
$\emph{\textbf{I}}_1$ & $\emph{\textbf{I}}_1$ &
$\emph{\textbf{I}}_0$ & $-\emph{\textbf{I}}_3$ &
$\emph{\textbf{I}}_2$  & $-\emph{\textbf{i}}_1$
& $-1$ & $-\emph{\textbf{i}}_3$ & $\emph{\textbf{i}}_2$  \\
$\emph{\textbf{I}}_2$ & $\emph{\textbf{I}}_2$ &
$\emph{\textbf{I}}_3$ & $\emph{\textbf{I}}_0$  &
$-\emph{\textbf{I}}_1$ & $-\emph{\textbf{i}}_2$
& $\emph{\textbf{i}}_3$  & $-1$ & $-\emph{\textbf{i}}_1$ \\
$\emph{\textbf{I}}_3$ & $\emph{\textbf{I}}_3$ &
$-\emph{\textbf{I}}_2$ & $\emph{\textbf{I}}_1$  &
$\emph{\textbf{I}}_0$  & $-\emph{\textbf{i}}_3$
& $-\emph{\textbf{i}}_2$ & $\emph{\textbf{i}}_1$  & $-1$ \\
\end{tabular}
\end{ruledtabular}
\end{table}

\subsection{Gravitational and electromagnetic fields}

The gravitational field and electromagnetic field both can be
illustrated by the quaternion, and their quaternion spaces will be
combined together to become the octonion space. In other words, the
characteristics of gravitational field and electromagnetic field can
be described with the octonion space simultaneously.

\subsubsection{Linear momentum}

In the quaternion space for the gravitational field, the basis
vector $\mathbb{E}_g$ = ($1$, $\emph{\textbf{i}}_1$,
$\emph{\textbf{i}}_2$, $\emph{\textbf{i}}_3$), and the radius vector
$\mathbb{R}_g$ = ($r_0$, $r_1$, $r_2$, $r_3$), with the velocity
$\mathbb{V}_g$ = ($v_0$, $v_1$, $v_2$, $v_3$). For the
electromagnetic field, the basis vector $\mathbb{E}_e$ =
($\emph{\textbf{I}}_0$, $\emph{\textbf{I}}_1$,
$\emph{\textbf{I}}_2$, $\emph{\textbf{I}}_3$), the radius vector
$\mathbb{R}_e$ = ($R_0$, $R_1$, $R_2$, $R_3$), and the velocity
$\mathbb{V}_e$ = ($V_0$, $V_1$, $V_2$, $V_3$), with $\mathbb{E}_e$ =
$\mathbb{E}_g$ $\circ$ $\emph{\textbf{I}}_0$ .

The $\mathbb{E}_e$ is independent of the $\mathbb{E}_g$ . Both of
them can be combined together to become the basis vector
$\mathbb{E}$ of the octonion space, that is,
\begin{eqnarray}
\mathbb{E} = (1, \emph{\textbf{i}}_1, \emph{\textbf{i}}_2,
\emph{\textbf{i}}_3, \emph{\textbf{I}}_0, \emph{\textbf{I}}_1,
\emph{\textbf{I}}_2, \emph{\textbf{I}}_3)~.
\end{eqnarray}

The octonion radius vector $\mathbb{R} = \mathbb{R}_g + k_{eg} \mathbb{R}_e$ in the octonion space is
\begin{eqnarray}
\mathbb{R} = && r_0 + \emph{\textbf{i}}_1 r_1 + \emph{\textbf{i}}_2
r_2 + \emph{\textbf{i}}_3 r_3
\nonumber\\
&& + k_{eg} (\emph{\textbf{I}}_0 R_0 + \emph{\textbf{I}}_1 R_1 +
\emph{\textbf{I}}_2 R_2 + \emph{\textbf{I}}_3 R_3)~,
\end{eqnarray}
and the octonion velocity $\mathbb{V} = \mathbb{V}_g + k_{eg} \mathbb{V}_e$ is
\begin{eqnarray}
\mathbb{V} = && v_0 + \emph{\textbf{i}}_1 v_1 + \emph{\textbf{i}}_2
v_2 + \emph{\textbf{i}}_3 v_3
\nonumber\\
&& + k_{eg} ( \emph{\textbf{I}}_0 V_0 + \emph{\textbf{I}}_1 V_1 +
\emph{\textbf{I}}_2 V_2 + \emph{\textbf{I}}_3 V_3)~,
\end{eqnarray}
where $r_0 = v_0 t$, $t$ is the time, and $v_0$ is the speed of
gravitational intermediate boson; the symbol $\circ$ denotes the
octonion multiplication.

When the electric charge is combined with the mass to become the
electron or the proton etc, we obtain the $R_i \emph{\textbf{I}}_i
= r_i \emph{\textbf{i}}_i \circ \emph{\textbf{I}}_0$ and $V_i
\emph{\textbf{I}}_i = v_i \emph{\textbf{i}}_i \circ
\emph{\textbf{I}}_0$ , with $\emph{\textbf{i}}_0 = 1$. In the same
way, the gravitational intermediate boson and electromagnetic
intermediate boson can be combined together to become the photon.

The potential of the gravitational and electromagnetic fields are
$\mathbb{A}_g = (a_0 , a_1 , a_2 , a_3)$ and $\mathbb{A}_e = (A_0 ,
A_1 , A_2 , A_3)$ respectively, with $\mathbb{A}_e = \mathbb{A}_q
\circ \emph{\textbf{I}}_0$. They can be combined together to become
the potential $\mathbb{A} = \mathbb{A}_g + k_{eg} \mathbb{A}_e $ .

The strength $\mathbb{B}(b_0, b_1, b_2, b_3, B_0, B_1, B_2, B_3)$
consists of gravitational strength $\mathbb{B}_g$ and
electromagnetic strength $\mathbb{B}_e$ . The gauge equations $b_0 =
0$ and $B_0 = 0$ . And
\begin{eqnarray}
\mathbb{B} = \lozenge \circ \mathbb{A} = \mathbb{B}_g + k_{eg} \mathbb{B}_e~.
\end{eqnarray}

The gravitational strength $\mathbb{B}_g$ in Eq.(2) includes two
components, $\textbf{g} = ( g_{01} , g_{02} , g_{03} ) $ and
$\textbf{b} = ( g_{23} , g_{31} , g_{12} )$, while the
electromagnetic strength $\mathbb{B}_e$ involves two parts,
$\textbf{E} = ( B_{01} , B_{02} , B_{03} ) $ and $\textbf{B} = (
B_{23} , B_{31} , B_{12} )$ .
\begin{eqnarray}
\textbf{E}/v_0 = && \emph{\textbf{I}}_1 ( \partial_0 A_1 +
\partial_1 A_0 ) + \emph{\textbf{I}}_2 ( \partial_0 A_2 + \partial_2
A_0 )
\nonumber\\
&& + \emph{\textbf{I}}_3 ( \partial_0 A_3 + \partial_3 A_0 )~,
\\
\textbf{B} = && \emph{\textbf{I}}_1 ( \partial_3 A_2 - \partial_2
A_3 ) + \emph{\textbf{I}}_2 ( \partial_1 A_3 - \partial_3 A_1 )
\nonumber\\
&& + \emph{\textbf{I}}_3 ( \partial_2 A_1 - \partial_1 A_2 )~.
\end{eqnarray}

\begin{table}[b]
\caption{\label{tab:table1}The operator and multiplication of the
physical quantity in the octonion space.}
\begin{ruledtabular}
\begin{tabular}{ll}
$definitions$               &  $meanings$                                            \\
\hline
$\nabla \cdot \textbf{a}$   &  $-(\partial_1 a_1 + \partial_2 a_2 + \partial_3 a_3)$  \\
$\nabla \times \textbf{a}$  &  $\emph{\textbf{i}}_1 ( \partial_2 a_3
                                 - \partial_3 a_2 ) + \emph{\textbf{i}}_2 ( \partial_3 a_1
                                 - \partial_1 a_3 )$                                  \\
$$                          &  $ + \emph{\textbf{i}}_3 ( \partial_1 a_2
                                 - \partial_2 a_1 )$                                  \\
$\nabla a_0$                &  $\emph{\textbf{i}}_1 \partial_1 a_0
                                 + \emph{\textbf{i}}_2 \partial_2 a_0
                                 + \emph{\textbf{i}}_3 \partial_3 a_0  $              \\
$\partial_0 \textbf{a}$     &  $\emph{\textbf{i}}_1 \partial_0 a_1
                                 + \emph{\textbf{i}}_2 \partial_0 a_2
                                 + \emph{\textbf{i}}_3 \partial_0 a_3  $              \\
\hline
$\nabla \cdot \textbf{A}$   &  $-(\partial_1 A_1 + \partial_2 A_2 + \partial_3 A_3) \emph{\textbf{I}}_0 $  \\
$\nabla \times \textbf{A}$  &  $-\emph{\textbf{I}}_1 ( \partial_2
                                 A_3 - \partial_3 A_2 ) - \emph{\textbf{I}}_2 ( \partial_3 A_1
                                 - \partial_1 A_3 )$                                 \\
$$                          &  $- \emph{\textbf{I}}_3 ( \partial_1 A_2 - \partial_2 A_1 )$    \\
$\nabla \circ \textbf{A}_0$ &  $\emph{\textbf{I}}_1 \partial_1 A_0
                                 + \emph{\textbf{I}}_2 \partial_2 A_0
                                 + \emph{\textbf{I}}_3 \partial_3 A_0  $             \\
$\partial_0 \textbf{A}$     &  $\emph{\textbf{I}}_1 \partial_0 A_1
                                 + \emph{\textbf{I}}_2 \partial_0 A_2
                                 + \emph{\textbf{I}}_3 \partial_0 A_3  $             \\
\end{tabular}
\end{ruledtabular}
\end{table}

In the octonion space, the electric current density $\mathbb{S}_e =
q \mathbb{V}_e$ is the source for the
electromagnetic field, and the linear momentum density $\mathbb{S}_g
= m \mathbb{V}_g $ for the gravitational field. The source
$\mathbb{S}$ satisfies
\begin{eqnarray}
\mu \mathbb{S} = && - ( \mathbb{B}/v_0 + \lozenge)^* \circ
\mathbb{B}
\nonumber\\
= && \mu_g \mathbb{S}_g + k_{eg} \mu_e \mathbb{S}_e - \mathbb{B}^*
\circ \mathbb{B}/v_0~,
\end{eqnarray}
where $k_{eg}^2 = \mu_g /\mu_e$, with $\mu_e = \mu_q$; $q$ is the electric charge
density; $*$ denotes the conjugate of octonion. While
\begin{eqnarray}
\mathbb{B}^* \circ \mathbb{B}/ \mu_g = \mathbb{B}_g^* \circ
\mathbb{B}_g / \mu_g + \mathbb{B}_e^* \circ \mathbb{B}_e / \mu_e~.
\end{eqnarray}

The force density $\mathbb{F}$ is defined from linear momentum
density $\mathbb{P} = \mu \mathbb{S} / \mu_g$ , which is the
extension of the $\mathbb{S}_g$ .
\begin{eqnarray}
\mathbb{F} = v_0 (\mathbb{B}/v_0 + \lozenge )^* \circ \mathbb{P}~,
\end{eqnarray}
where the force density includes Lorentz force density, gravity
density, and inertial force density.

\begin{table}[t]
\caption{\label{tab:table1}The electromagnetic force and gravity in
the quaternion space $\mathbb{E}_g$ of octonion space.}
\begin{ruledtabular}
\begin{tabular}{ll}
$ definitions $                                 & $ meanings $ \\
\hline
$v_0 \partial_0 \textbf{p}$                     & inertial force density\\
$m \textbf{g}^*$                                & gravity density  \\
$q \textbf{E}^* \circ \emph{\textbf{I}}_0$      & electric force density\\
$q \textbf{B}^* \times \textbf{V}$              & magnetic force density\\
$q v_0 \textbf{B}^* \circ \emph{\textbf{I}}_0$  & new force part \\
$q \textbf{E}^* \times \textbf{V}/v_0$          & new force part \\
$\triangle m \textbf{g}^*$                      & new force part \\
$p_0 \textbf{b}^* $                             & new force part \\
$v_0 \nabla^* p_0$                              & new force part \\
$\textbf{g}^* \times \textbf{p}/v_0$            & new force part \\
$\textbf{b}^* \times \textbf{p}$                & new force part \\
$v_0 \nabla^* \times \textbf{p}$                & new force part \\
\end{tabular}
\end{ruledtabular}
\end{table}

\subsubsection{Electromagnetic force and gravity}

In the octonion space, the gravitational mass density $\widehat{m} =
m + \triangle m$, with $\triangle m = - \mathbb{B}^* \circ
\mathbb{B} / (\mu_g v_0^2)$. The linear momentum density $\mathbb{P}
= p_0 + \Sigma (p_j \emph{\textbf{i}}_j ) + \Sigma (P_i
\emph{\textbf{I}}_i ) $. And $m$ is the inertial mass density; $P_i
= M V_i $, $M = k_{eg} \mu_e q / \mu_g$; $p_0 = \widehat{m} v_0$,
$p_j = m v_j $; $\textbf{v} = \Sigma (v_j \emph{\textbf{i}}_j) $,
$\textbf{V} = \Sigma (V_j \emph{\textbf{I}}_j) $.

By Eq.(25), the force density $\mathbb{F}$ is
\begin{eqnarray}
\mathbb{F} = f_0 + \Sigma (f_j \emph{\textbf{i}}_j ) + \Sigma (F_i
\emph{\textbf{I}}_i)~,
\end{eqnarray}
where $f_0 = \partial p_0 / \partial t + v_0 \Sigma (  \partial p_j
/ \partial r_j ) + \Sigma ( b_j p_j + B_j P_j ) $.

The vectorial part $\textbf{f} = \Sigma (f_j \emph{\textbf{i}}_j )$
and  $\textbf{F} = \Sigma (F_i \emph{\textbf{I}}_i )$ of force
density $\mathbb{F}$ can be decomposed from Eq.(26).
\begin{eqnarray}
\textbf{f} = && v_0 \partial_0 \textbf{p} + (\textbf{g}/v_0 +
\textbf{b} + v_0 \nabla)^* \times \textbf{p}
\nonumber\\
&& + q (\textbf{E}/v_0 + \textbf{B})^* \times \textbf{V} + p_0
(\textbf{g}/v_0 + \textbf{b})^*
\nonumber\\
&& + v_0 \nabla^* p_0 + q v_0 (\textbf{E}/v_0 + \textbf{B})^* \circ
\emph{\textbf{I}}_0~,
\\
\textbf{F} = && v_0 \partial_0 \textbf{P} + ( \textbf{g}/v_0 +
\textbf{b} + v_0 \nabla)^* \times \textbf{P}
\nonumber\\
&& + k_{eg} (\textbf{E}/v_0 + \textbf{B})^* \circ \emph{\textbf{p}}
+ k_{eg} p_0 (\textbf{E}/v_0 + \textbf{B})^*
\nonumber\\
&& + v_0 \lozenge^* \circ \textbf{P}_0 + (\textbf{g}/v_0 +
\textbf{b})^* \circ \textbf{P}_0~,
\end{eqnarray}
where $\textbf{P} = \Sigma (P_j \emph{\textbf{I}}_j)$ ,
$\textbf{P}_0 = P_0 \emph{\textbf{I}}_0$ .

In the paper, the physical quantity in the space $\mathbb{E}_g$ can
be detected, but that in $\mathbb{E}_e$ can not yet.

In the physics world, the term $\textbf{f}$ can be detected, which
includes the inertial force, electromagnetic force, gravity and some
other new kinds of forces. But the term $\textbf{F}$ may not be
detected at present. Meanwhile we can obtain the mass continuity equation from $f_0 = 0$, and the charge continuity equation from $F_0 = 0$ respectively.

Comparing Table IV with Table VII, we find that some terms of forces
in the former table can not be detected in the physics space
$\mathbb{E}_g$. The above means that the quaternion space
$\mathbb{E}_g$ or $\mathbb{E}_q$ is not suitable for describing the
electromagnetic force in the physics world.

\section{Forces from energies}

More force terms can be defined from the energy in the electromagnetic
and gravitational fields. Defining from the linear momentum, the force definition can not cover the gradient of
energy sufficiently, so that we need one new definition for the force. The new
definition of forces is defined from the energy. Those new forces
include the inertial force, gravitational force, electric force,
magnetic force, gradient of energy, and interacting force between
the magnetic strength with magnetic moment, etc.

\subsection{Gravitational field}

To incorporate various kinds of energies within a single definition,
the angular momentum and energy will both be extended to apply
within gravitational fields.

The angular momentum density $\mathbb{L}_g = l_0 + \Sigma (l_j
\emph{\textbf{i}}_j )$ is defined from the linear momentum density
$\mathbb{P}_g$ and the radius vector $\mathbb{R}_g$ , and can be rewritten as follows
\begin{eqnarray}
\mathbb{L}_g = (\mathbb{R}_g + k_{rx} \mathbb{X}_g ) \circ
\mathbb{P}_g~,
\end{eqnarray}
with
\begin{eqnarray}
l_0 = && (r_0 + k_{rx} x_0) p_0 + (\textbf{r} + k_{rx} \textbf{x})
\cdot \textbf{p}~,
\\
\textbf{l} = && (r_0 + k_{rx} x_0) \textbf{p} + p_0 (\textbf{r} +
k_{rx} \textbf{x})
\nonumber\\
&& + (\textbf{r} + k_{rx} \textbf{x}) \times \textbf{p}~,
\end{eqnarray}
where $\textbf{l} = \Sigma (l_j \emph{\textbf{i}}_j )$; $k_{rx}$ is
a coefficient. The quaternion quantity $\mathbb{X}_g = \Sigma (x_i
\emph{\textbf{i}}_i)$ is similar to Hertz vector in the electrodynamics
theory. The derivation of $\mathbb{X}_g$ will yield the gravitational potential, with $k_{rx}\mathbb{X}_g \ll \mathbb{R}_g$ .

The angular momentum density $\mathbb{L}_g$ includes the scalar part $l_0$, orbital
angular momentum density $\textbf{r} \times \textbf{p}$, and some other omissible terms etc.

\subsubsection{Energy and torque}

We choose the following definition of energy to include various
energies in the gravitational field. In quaternion space, the
quaternion energy density $\mathbb{W}_g = w_0 + \Sigma (w_j
\emph{\textbf{i}}_j )$ is defined from the angular momentum density
$\mathbb{L}_g$ .
\begin{eqnarray}
\mathbb{W}_g = v_0 ( \mathbb{B}_g/v_0 + \lozenge) \circ \mathbb{L}_g~,
\end{eqnarray}
where the $-w_0/2$ is the energy density, which includes the kinetic
energy, potential energy, field energy, and work etc; the
$\textbf{w}/2 = \Sigma (w_j \emph{\textbf{i}}_j )/2$ is the torque.

Expressing the energy density as
\begin{eqnarray}
w_0 = && v_0 \partial_0 l_0 + v_0 \nabla \cdot \textbf{l} +
(\textbf{g} / v_0 + \textbf{b}) \cdot \textbf{l}
\nonumber\\
= &&
(\textbf{g}/v_0 + \textbf{b}) \cdot (r_0 \textbf{p} + p_0 \textbf{r}
+ \textbf{r} \times \textbf{p} ) \nonumber
\\
&& + k_{rx} (\textbf{g}/v_0 + \textbf{b}) \cdot (x_0 \textbf{p} +
p_0 \textbf{x} + \textbf{x} \times \textbf{p} ) \nonumber
\\
&& + v_0 \nabla \cdot (r_0 \textbf{p} + p_0 \textbf{r} + \textbf{r}
\times \textbf{p} ) \nonumber
\\
&& + v_0 \partial_0 (r_0 p_0 + \textbf{r} \cdot \textbf{p}) + v_0
k_{rx} \partial_0 (x_0 p_0 + \textbf{x} \cdot \textbf{p}) \nonumber
\\
&& + v_0 k_{rx} \nabla \cdot (x_0 \textbf{p} + p_0 \textbf{x} +
\textbf{x} \times \textbf{p} )~,
\end{eqnarray}
where the term $ (\textbf{g}/v_0 + \textbf{b}) \cdot (x_0
\textbf{p} + p_0 \textbf{x} + \textbf{x} \times \textbf{p})$ is one
new kind of energy; $\textbf{x} = \Sigma (x_j \emph{\textbf{i}}_j
)$.

In case of $r_0 = 0$, $x_0 = 0$, $\textbf{b} = 0$,
$\textbf{r} \parallel \textbf{p}$, and $\textbf{x}
\parallel \textbf{p}$, the above can be reduced to
\begin{eqnarray}
- w_0/2 \approx && m v_0^2 + \triangle m_1 v_0^2
- \textbf{a} \cdot \textbf{p}/2 - a_0 p_0/2
\nonumber
\\
&& - v_0 \partial_0 ( \textbf{r} \cdot \textbf{p})/2
- (\textbf{g}/v_0) \cdot (p_0 \textbf{r})/2 ~,
\nonumber
\end{eqnarray}
where $a_0 / v_0 = \partial_0 x_0 + \nabla \cdot \textbf{x}$ and $\textbf{a} = \partial_0 \textbf{x} + \nabla x_0 + \nabla \times \textbf{x}$ are the scalar and vectorial potential of gravitational field respectively; the last three terms in the above are equal to the sum of kinetic energy and gravitational potential energy; comparing with the potential energy in classical field theory, we choose $k_{rx} = 1 / v_0$ in the paper.

In a similar way, expressing the torque density as
\begin{eqnarray}
\textbf{w} = && v_0 \partial_0 \textbf{l} + v_0 \nabla l_0 + v_0
\nabla \times \textbf{l}
\nonumber\\
&& + l_0 (\textbf{g} / v_0 + \textbf{b}) + (\textbf{g} / v_0 +
\textbf{b}) \times \textbf{l}
\nonumber\\
= && (\textbf{g}/v_0 + \textbf{b}) \times (r_0 \textbf{p}
+ p_0 \textbf{r} + \textbf{r} \times \textbf{p} ) \nonumber
\\
&& + (r_0 p_0 + \textbf{r} \cdot \textbf{p}) (\textbf{g}/v_0 +
\textbf{b}) \nonumber
\\
&& + k_{rx} (\textbf{g}/v_0 + \textbf{b}) \times (x_0 \textbf{p} +
p_0 \textbf{x} + \textbf{x} \times \textbf{p} ) \nonumber
\\
&& + k_{rx} (x_0 p_0 + \textbf{x} \cdot \textbf{p}) (\textbf{g}/v_0
+ \textbf{b}) \nonumber
\\
&& + v_0 \nabla \times (r_0 \textbf{p} + p_0 \textbf{r} + \textbf{r}
\times \textbf{p} ) \nonumber
\\
&& + v_0 \nabla (r_0 p_0 + \textbf{r} \cdot \textbf{p}) + v_0
\partial_0 (r_0 \textbf{p} + p_0 \textbf{r} + \textbf{r} \times
\textbf{p} ) \nonumber
\\
&& + v_0 k_{rx} \nabla \times (x_0 \textbf{p} + p_0 \textbf{x} +
 \textbf{x} \times \textbf{p} ) \nonumber
\\
&& + v_0 k_{rx} \nabla (x_0 p_0 + \textbf{x} \cdot \textbf{p})
\nonumber
\\
&& + v_0 k_{rx} \partial_0 (x_0 \textbf{p} + p_0 \textbf{x} +
\textbf{x} \times \textbf{p})~.
\end{eqnarray}

In case of $r_0 = 0$, $x_0 = 0$, $ (\textbf{v}^* \cdot \textbf{v}) \ll v_0^2 $, and
$\textbf{b} = 0$, the above can be reduced to
\begin{eqnarray}
\textbf{w} \approx
&& (\textbf{g}/v_0) \times ( p_0 \textbf{r} ) + v_0 \partial_0 ( \textbf{r} \times \textbf{p} + p_0 \textbf{r} )
\nonumber
\\
&& - 2 v_0 \textbf{p} + v_0 \textbf{r} ( \nabla \cdot \textbf{p} )
+ v_0 \textbf{r} \times ( \nabla \times \textbf{p} )
\nonumber
\\
&& + a_0 \textbf{p} + p_0 \textbf{a} + \textbf{a} \times \textbf{p} ~,
\nonumber
\end{eqnarray}
where the first and second terms are the torque caused by gravity and some other force terms respectively.

\begin{table}[h]
\caption{\label{tab:table1}Some related physical quantities in the quaternion space.}
\begin{ruledtabular}
\begin{tabular}{ll}
$ definitions $           & $ meanings $ \\
\hline
$p_0$                     & scalar part \\
$\textbf{p}$              & linear momentum density \\
$l_0$                     & scalar part \\
$\textbf{l}$              & angular momentum density \\
$w_0$                     & energy density \\
$\textbf{w}$              & torque density \\
$n_0$                     & power density \\
$\textbf{n}$              & vectorial part \\
\end{tabular}
\end{ruledtabular}
\end{table}

\subsubsection{Power and force}

In the quaternion space, the quaternion power density $\mathbb{N}_g
= n_0 + \Sigma (n_j \emph{\textbf{i}}_j )$ is defined from the
$\mathbb{W}_g$ ,
\begin{eqnarray}
\mathbb{N}_g = v_0 ( \mathbb{B}_g/v_0 + \lozenge)^* \circ
\mathbb{W}_g~,
\end{eqnarray}
where the $f_0 = -n_0/(2 v_0)$ is the power density, while the
vectorial part $\textbf{n} = \Sigma (n_j \emph{\textbf{i}}_j )$ is
the function of forces.

Expressing the scalar $n_0$ as
\begin{eqnarray}
n_0 =  v_0 \partial_0 w_0 + v_0 \nabla^* \cdot \textbf{w} +
(\textbf{g}/v_0 + \textbf{b})^* \cdot \textbf{w}~.
\end{eqnarray}

In a similar way, expressing the vectorial part $\textbf{n}$ of
$\mathbb{N}_g$ as follows
\begin{eqnarray}
\textbf{n} = && v_0 \nabla^* w_0  + v_0 \partial_0 \textbf{w} + v_0
\nabla^* \times \textbf{w}
\nonumber \\
&& + (\textbf{g}/v_0 + \textbf{b})^* \times \textbf{w} + w_0
(\textbf{g}/v_0 + \textbf{b})^*~.
\nonumber
\end{eqnarray}

The force $\textbf{f} = - \textbf{n} /(2 v_0)$ in the gravitational
field can be defined from the vectorial part $\textbf{n}$ .
\begin{eqnarray}
- 2 \textbf{f} = && \nabla^* w_0 + \partial_0 \textbf{w} +
(\textbf{g}/v_0 + \textbf{b})^* \times \textbf{w}/v_0
\nonumber \\
&& + \nabla^* \times \textbf{w} + w_0 (\textbf{g}/v_0 +
\textbf{b})^*/v_0~,
\end{eqnarray}
where the force $\textbf{f}$ includes the gravity, inertial force,
and gradient of energy etc.

In the gravitational field, the above means that Eq.(37) encompasses
the force terms in Eq.(8) and some other terms of the gradient of energy
etc.

In case of $\mathbb{N}_g = 0$ in Eq.(35), we can obtain the mass continuity equation from $n_0 = 0$ in Eq.(36), as well as the force equilibrium equation from $\textbf{n} = 0$ in Eq.(37).

\subsection{Gravitational and electromagnetic fields}

In the case for coexistence of electromagnetic field and
gravitational field, the octonionic angular momentum density
$\mathbb{L} = l_0 + \Sigma (l_j \emph{\textbf{i}}_j ) + L_0
\emph{\textbf{I}}_0 + \Sigma (L_j \emph{\textbf{I}}_j)$ is defined
from the octonion linear momentum density $\mathbb{P}$ and octonion
radius vector $\mathbb{R}$ in Eqs.(18) and (19).
\begin{eqnarray}
\mathbb{L} = (\mathbb{R} + k_{rx} \mathbb{X} ) \circ \mathbb{P}~,
\end{eqnarray}
with
\begin{eqnarray}
l_0 = && (r_0 + k_{rx} x_0) p_0 + (\textbf{R}_0 + k_{rx}
\textbf{X}'_0) \circ \textbf{P}_0
\nonumber\\
&& + (\textbf{r} + k_{rx} \textbf{x}) \cdot \textbf{p} + (\textbf{R}
+ k_{rx} \textbf{X}') \cdot \textbf{P}~,
\\
\textbf{l} = && (r_0 + k_{rx} x_0) \textbf{p} + (\textbf{R} + k_{rx}
\textbf{X}') \circ \textbf{P}_0
\nonumber\\
&& + (\textbf{r} + k_{rx} \textbf{x}) \times \textbf{p} +
(\textbf{R}_0 + k_{rx} \textbf{X}'_0) \circ \textbf{P}
\nonumber\\
&& + p_0 (\textbf{r} + k_{rx} \textbf{x}) + (\textbf{R} + k_{rx}
\textbf{X}') \times \textbf{P}~,
\\
\textbf{L}_0 = && (r_0 + k_{rx} x_0) \textbf{P}_0 + (\textbf{r} +
k_{rx} \textbf{x}) \cdot \textbf{P}
\nonumber\\
&& + p_0 (\textbf{R}_0 + k_{rx} \textbf{X}'_0) + (\textbf{R} + k_{rx}
\textbf{X}') \cdot \textbf{p}~,
\\
\textbf{L} = && (r_0 + k_{rx} x_0) \textbf{P} + (\textbf{r} + k_{rx}
\textbf{x}) \circ \textbf{P}_0
\nonumber\\
&& + (\textbf{r} + k_{rx} \textbf{x}) \times \textbf{P} + (\textbf{R}_0 + k_{rx} \textbf{X}'_0) \circ \textbf{p}
\nonumber\\
&& + p_0 (\textbf{R} + k_{rx} \textbf{X}') + (\textbf{R} + k_{rx} \textbf{X}') \times \textbf{p}~,
\end{eqnarray}
where $\mathbb{X} = \Sigma (x_i \emph{\textbf{i}}_i) + k_{eg} \Sigma (X_i \emph{\textbf{I}}_i)$; $\textbf{X}'_0 = k_{eg} X_0 \emph{\textbf{I}}_0 $;
$\textbf{X}' = k_{eg} \Sigma (X_j \emph{\textbf{I}}_j )$. $\textbf{L}_0 = L_0 \emph{\textbf{I}}_0$; $\textbf{L} = \Sigma (L_j
\emph{\textbf{I}}_j)$. $\textbf{P}_0 = P_0 \emph{\textbf{I}}_0 $; $\textbf{P} = \Sigma (P_j \emph{\textbf{I}}_j )$. $\textbf{R}_0 =
R_0 \emph{\textbf{I}}_0 $; $\textbf{R} = \Sigma (R_j \emph{\textbf{I}}_j )$. Similarly the derivation of octonion
physics quantity $\mathbb{X}$ will yield the gravitational potential and electromagnetic potential.

\begin{table}[t]
\caption{\label{tab:table1}Some physical quantity in the quaternion space and octonion space.}
\begin{ruledtabular}
\begin{tabular}{ll}
$ definitions $                                       & $ meanings $ \\
\hline
$\mathbb{X}$                                          & field quantity \\
$\mathbb{A} = \lozenge \circ \mathbb{X}$              & field potential \\
$\mathbb{B} = \lozenge \circ \mathbb{A}$              & field strength \\
$\mathbb{R}$                                          & radius vector \\
$\mathbb{V} = v_0 \lozenge \circ \mathbb{R}$          & velocity \\
$\mathbb{U} = \lozenge \circ \mathbb{V}$              & velocity curl \\
$\mu \mathbb{S} = - ( \mathbb{B} /v_0 + \lozenge )^* \circ \mathbb{B}$              & field source \\
$\mathbb{P} = \mu \mathbb{S} / \mu_g$                 & linear momentum density \\
$\mathbb{R}' = \mathbb{R} + k_{rx} \mathbb{X}$        & compounding radius vector \\
$\mathbb{L} = \mathbb{R}' \circ \mathbb{P}$           & angular momentum density \\
$\mathbb{W} = v_0 ( \mathbb{B} /v_0 + \lozenge ) \circ \mathbb{L}$                  & energy and torque densities \\
$\mathbb{N} = v_0 ( \mathbb{B} /v_0 + \lozenge )^* \circ \mathbb{W}$                & power density \\
$\mathbb{F} = - \mathbb{N} / (2v_0)$                  & force density \\
\end{tabular}
\end{ruledtabular}
\end{table}

\subsubsection{Energy and torque}

In the case for coexistence of electromagnetic field and
gravitational field, the octonion energy density $\mathbb{W} = w_0 +
\Sigma (w_j \emph{\textbf{i}}_j ) + \Sigma (W_i \emph{\textbf{I}}_i
)$ is defined from the octonion angular momentum density
$\mathbb{L}$ and octonion field strength $\mathbb{B}$ in Eqs.(20)
and (38).
\begin{eqnarray}
\mathbb{W} = v_0 ( \mathbb{B}/v_0 + \lozenge) \circ \mathbb{L}~,
\end{eqnarray}
where the $-w_0/2$ is the energy density, which includes the kinetic
energy, potential energy, field energy, and work etc; the
$\textbf{w}/2 = \Sigma (w_j \emph{\textbf{i}}_j )/2$ is the torque.

Expressing the energy density as
\begin{eqnarray}
w_0 = && v_0 \partial_0 l_0 + v_0 \nabla \cdot \textbf{l} +
(\textbf{g} / v_0 + \textbf{b}) \cdot \textbf{l}
\nonumber\\
&& + k_{eg} (\textbf{E} / v_0 + \textbf{B}) \cdot \textbf{L}
\nonumber
\\
= && v_0 \partial_0 [(r_0 + k_{rx} x_0) p_0 + (\textbf{R}_0 + k_{rx}
\textbf{X}'_0) \circ \textbf{P}_0]
\nonumber
\\
&& + v_0 \partial_0 [(\textbf{r} + k_{rx} \textbf{x}) \cdot
\textbf{p} + (\textbf{R} + k_{rx} \textbf{X}') \cdot \textbf{P}]
\nonumber\\
&& + (v_0 \nabla + \textbf{g} / v_0 + \textbf{b}) \cdot [(r_0 +
k_{rx} x_0) \textbf{p}]
\nonumber\\
&& + (v_0 \nabla + \textbf{g} / v_0 + \textbf{b}) \cdot [(\textbf{R}
+ k_{rx} \textbf{X}') \circ \textbf{P}_0]
\nonumber\\
&& + (v_0 \nabla + \textbf{g} / v_0 + \textbf{b}) \cdot [
(\textbf{r} + k_{rx} \textbf{x}) \times \textbf{p}]
\nonumber\\
&& + (v_0 \nabla + \textbf{g} / v_0 + \textbf{b}) \cdot [
(\textbf{R}_0 + k_{rx} \textbf{X}'_0) \circ \textbf{P}]
\nonumber\\
&& + (v_0 \nabla + \textbf{g} / v_0 + \textbf{b}) \cdot [p_0
(\textbf{r} + k_{rx} \textbf{x})]
\nonumber\\
&& + (v_0 \nabla + \textbf{g} / v_0 + \textbf{b}) \cdot [(\textbf{R}
+ k_{rx} \textbf{X}') \times \textbf{P}]
\nonumber\\
&& + k_{eg} (\textbf{E} / v_0 + \textbf{B}) \cdot [(r_0 + k_{rx} x_0) \textbf{P}]
\nonumber\\
&& + k_{eg} (\textbf{E} / v_0 + \textbf{B}) \cdot [(\textbf{r} + k_{rx} \textbf{x}) \circ \textbf{P}_0]
\nonumber\\
&& + k_{eg} (\textbf{E} / v_0 + \textbf{B}) \cdot [(\textbf{r} + k_{rx} \textbf{x}) \times \textbf{P}]
\nonumber\\
&& + k_{eg} (\textbf{E} / v_0 + \textbf{B}) \cdot [ (\textbf{R}_0 + k_{rx} \textbf{X}'_0) \circ \textbf{p}]
\nonumber\\
&& + k_{eg} (\textbf{E} / v_0 + \textbf{B}) \cdot [  p_0 (\textbf{R} + k_{rx} \textbf{X}')]
\nonumber\\
&& + k_{eg} (\textbf{E} / v_0 + \textbf{B}) \cdot [ (\textbf{R} + k_{rx} \textbf{X}') \times \textbf{p} ]
~,
\end{eqnarray}
where the $-w_0/2$ includes some new kinds of energies.

In case of $r_0 = 0$, $x_0 = 0$, $\textbf{b} = 0$,
$\textbf{r} \parallel \textbf{p}$, and $\textbf{x}
\parallel \textbf{p}$, the above can be reduced to
\begin{eqnarray}
- w_0/2 \approx  && m v_0^2 + \triangle m v_0^2 - a_0 p_0/2 - \textbf{a} \cdot \textbf{p}/2
\nonumber\\
&& - v_0 \partial_0 (\textbf{r} \cdot \textbf{p})/2 -
(\textbf{g}/v_0) \cdot (p_0 \textbf{r})/2
\nonumber\\
&& - k_{eg} \textbf{A}_0 \circ \textbf{P}_0 / 2 - k_{eg} (\textbf{E}/v_0) \cdot (\textbf{r} \circ \textbf{P}_0)/2
\nonumber\\
&& - k_{eg} \textbf{A} \cdot \textbf{P} / 2 - k_{eg} \textbf{B} \cdot (\textbf{r} \times \textbf{P})/2~,
\nonumber
\end{eqnarray}
where $\textbf{A}_0 / v_0 = \partial_0 \textbf{X}_0 + \nabla \cdot
\textbf{X}$ and $\textbf{A} = \partial_0 \textbf{X} + \nabla
\circ \textbf{X}_0 + \nabla \times \textbf{X}$ are the
scalar and vectorial potential of electromagnetic field
respectively; the last four terms are the electric potential energy,
magnetic potential energy, and interacting energy between dipole
moment with electromagnetic strength in classical field theory.

In the electromagnetic field and gravitational field, the torque
$\textbf{w}/2$ in the space $\mathbb{E}_g$ can be detected, but the
other vectorial part $\Sigma (W_i \emph{\textbf{I}}_i)$ can not
currently.

In a similar way, expressing the torque density $\textbf{w}$ as
\begin{eqnarray}
\textbf{w} = && v_0 \partial_0 \textbf{l} + v_0 \nabla l_0 + v_0
\nabla \times \textbf{l}
\nonumber\\
&& + l_0 (\textbf{g} / v_0 + \textbf{b}) + (\textbf{g} / v_0 +
\textbf{b}) \times \textbf{l}
\nonumber\\
&& + k_{eg} (\textbf{E} / v_0 + \textbf{B}) \times \textbf{L}
\nonumber\\
&& + k_{eg} (\textbf{E} / v_0 + \textbf{B}) \circ \textbf{L}_0
\nonumber\\
= && v_0 \partial_0  [(r_0 + k_{rx} x_0) \textbf{p} + (\textbf{R} +
k_{rx} \textbf{X}') \circ \textbf{P}_0]
\nonumber\\
&& + v_0 \partial_0  [ (\textbf{r} + k_{rx} \textbf{x}) \times
\textbf{p} + (\textbf{R}_0 + k_{rx} \textbf{X}'_0) \circ \textbf{P}]
\nonumber\\
&& + v_0 \partial_0  [p_0 (\textbf{r} + k_{rx} \textbf{x}) +
(\textbf{R} + k_{rx} \textbf{X}') \times \textbf{P}]
\nonumber\\
&& + v_0 \nabla [(r_0 + k_{rx} x_0) p_0 + (\textbf{R}_0 + k_{rx}
\textbf{X}'_0) \circ \textbf{P}_0]
\nonumber\\
&& + v_0 \nabla [(\textbf{r} + k_{rx} \textbf{x}) \cdot \textbf{p} +
(\textbf{R} + k_{rx} \textbf{X}') \cdot \textbf{P}]
\nonumber\\
&& + [(r_0 + k_{rx} x_0) p_0 + (\textbf{r} + k_{rx} \textbf{x}) \cdot \textbf{p}] (\textbf{g} / v_0 + \textbf{b})
\nonumber\\
&& + [(\textbf{R}_0 + k_{rx} \textbf{X}'_0) \circ \textbf{P}_0]
(\textbf{g} / v_0 + \textbf{b})
\nonumber\\
&& + [(\textbf{R} + k_{rx} \textbf{X}') \cdot \textbf{P}] (\textbf{g} / v_0
+ \textbf{b})
\nonumber\\
&& + (v_0 \nabla + \textbf{g} / v_0 + \textbf{b}) \times [(r_0 +
k_{rx} x_0) \textbf{p}]
\nonumber\\
&& + (v_0 \nabla + \textbf{g} / v_0 + \textbf{b}) \times
[(\textbf{R} + k_{rx} \textbf{X}') \circ \textbf{P}_0]
\nonumber\\
&& + (v_0 \nabla + \textbf{g} / v_0 + \textbf{b}) \times [
(\textbf{r} + k_{rx} \textbf{x}) \times \textbf{p}]
\nonumber\\
&& + (v_0 \nabla + \textbf{g} / v_0 + \textbf{b}) \times [
(\textbf{R}_0 + k_{rx} \textbf{X}'_0) \circ \textbf{P}]
\nonumber\\
&& + (v_0 \nabla + \textbf{g} / v_0 + \textbf{b}) \times [p_0
(\textbf{r} + k_{rx} \textbf{x})]
\nonumber\\
&& + (v_0 \nabla + \textbf{g} / v_0 + \textbf{b}) \times
[(\textbf{R} + k_{rx} \textbf{X}') \times \textbf{P}]
\nonumber\\
&& + k_{eg} (\textbf{E} / v_0 + \textbf{B}) \times [(r_0 + k_{rx} x_0) \textbf{P}]
\nonumber\\
&& + k_{eg} (\textbf{E} / v_0 + \textbf{B}) \times [(\textbf{r} + k_{rx} \textbf{x}) \circ \textbf{P}_0]
\nonumber\\
&& + k_{eg} (\textbf{E} / v_0 + \textbf{B}) \times [(\textbf{r} + k_{rx} \textbf{x}) \times \textbf{P}]
\nonumber\\
&& + k_{eg} (\textbf{E} / v_0 + \textbf{B}) \times [ (\textbf{R}_0 + k_{rx} \textbf{X}'_0) \circ \textbf{p}]
\nonumber\\
&& + k_{eg} (\textbf{E} / v_0 + \textbf{B}) \times [  p_0 (\textbf{R} + k_{rx} \textbf{X}')]
\nonumber\\
&& + k_{eg} (\textbf{E} / v_0 + \textbf{B}) \times [ (\textbf{R} + k_{rx} \textbf{X}') \times \textbf{p} ]
\nonumber\\
&& + k_{eg} (\textbf{E} / v_0 + \textbf{B}) \circ [ (r_0 + k_{rx} x_0) \textbf{P}_0 ]
\nonumber\\
&& + k_{eg} (\textbf{E} / v_0 + \textbf{B}) \circ [ (\textbf{r} + k_{rx} \textbf{x}) \cdot \textbf{P} ]
\nonumber\\
&& + k_{eg} (\textbf{E} / v_0 + \textbf{B}) \circ [ p_0 (\textbf{R}_0 + k_{rx} \textbf{X}'_0)  ]
\nonumber\\
&& + k_{eg} (\textbf{E} / v_0 + \textbf{B}) \circ [ (\textbf{R} + k_{rx} \textbf{X}') \cdot \textbf{p}  ]
~.
\end{eqnarray}

In case of $r_0 = 0$, $x_0 = 0$, $ (\textbf{v}^* \cdot \textbf{v}) \ll v_0^2 $, and
$\textbf{b} = 0$, the above can be reduced to
\begin{eqnarray}
\textbf{w} \approx && (\textbf{g}/v_0) \times ( p_0 \textbf{r} ) +
v_0 \partial_0 ( \textbf{r} \times \textbf{p}) +  k_{eg} \textbf{A}_0 \circ \textbf{P}
\nonumber\\
&& + \textbf{a} \times \textbf{p} + a_0 \textbf{p} + p_0 \textbf{a} +  k_{eg} \textbf{A} \times \textbf{P} - 2 v_0 \textbf{p}
\nonumber\\
&& +  k_{eg} \textbf{A} \circ \textbf{P}_0 + v_0 \textbf{r} ( \nabla \cdot \textbf{p} ) + v_0 \textbf{r} \times ( \nabla \times \textbf{p} )
\nonumber\\
&& + k_{eg} (\textbf{E}/v_0) \times (\textbf{r} \circ \textbf{P}_0) + k_{eg}
\textbf{B} \times (\textbf{r} \times \textbf{P})~,
\nonumber
\end{eqnarray}
where the above includes the torque caused by gravity,
electromagnetic force, and other force terms etc.

Other parts of torque can be rewritten as follows
\begin{eqnarray}
\textbf{W}_0 = && v_0 \partial_0 \textbf{L}_0 + k_{eg} (\textbf{E} /
v_0 + \textbf{B}) \cdot \textbf{l}
\nonumber\\
&& + v_0 \nabla \cdot \textbf{L} + (\textbf{g} / v_0 + \textbf{b})
\cdot \textbf{L}~,
\nonumber\\
\textbf{W} = && v_0 \partial_0 \textbf{L} + l_0 k_{eg} (\textbf{E} /
v_0 + \textbf{B}) + v_0 \nabla \times \textbf{L}
\nonumber\\
&& + v_0 \nabla \circ \textbf{L}_0 + k_{eg} (\textbf{E} / v_0 +
\textbf{B}) \times \textbf{l}
\nonumber\\
&& + (\textbf{g} / v_0 + \textbf{b}) \times \textbf{L} + (\textbf{g}
/ v_0 + \textbf{b}) \circ \textbf{L}_0~,
\nonumber
\end{eqnarray}
where $\textbf{W}_0 = W_0 \emph{\textbf{I}}_0$, $\textbf{W} =
\Sigma (W_j \emph{\textbf{I}}_j)$.

The torque part $\Sigma (W_i \emph{\textbf{I}}_i)$ can not be
detected in $\mathbb{E}_g$, but it has still an effect on the power
and forces.

\subsubsection{Power and force}

In the octonion space, the octonion power density $\mathbb{N} = n_0
+ \Sigma (n_j \emph{\textbf{i}}_j) + \Sigma (N_i
\emph{\textbf{I}}_i)$ is defined from the octonion energy density
$\mathbb{W}$ and field strength $\mathbb{B}$ .
\begin{eqnarray}
\mathbb{N} = v_0 ( \mathbb{B}/v_0 + \lozenge)^* \circ \mathbb{W}~,
\end{eqnarray}
where $f_0 = - n_0/(2 v_0)$ is the power density, while one
vectorial part $\textbf{n} = \Sigma (n_j \emph{\textbf{i}}_j )$ is
the function of forces. But the other vectorial parts
$\textbf{N}_0 = N_0 \emph{\textbf{I}}_0$ and $\textbf{N} = \Sigma (N_j \emph{\textbf{I}}_j)$
may not be detected in the space
$\mathbb{E}_g$ at present.

Further expressing the scalar $n_0$ as
\begin{eqnarray}
n_0 = && v_0 \partial_0 w_0 + v_0 \nabla^* \cdot \textbf{w} +
(\textbf{g}/v_0 + \textbf{b})^* \cdot \textbf{w}
\nonumber\\
&& + k_{eg} (\textbf{E}/v_0 + \textbf{B})^* \cdot \textbf{W}~.
\end{eqnarray}

In a similar way, expressing the vectorial part $\textbf{n}$ of
$\mathbb{N}$ as follows
\begin{eqnarray}
\textbf{n} = && v_0 \nabla^* w_0  + v_0 \partial_0 \textbf{w} + v_0
\nabla^* \times \textbf{w}
\nonumber \\
&& + (\textbf{g}/v_0 + \textbf{b})^* \times \textbf{w} + w_0
(\textbf{g}/v_0 + \textbf{b})^*
\nonumber\\
&& + k_{eg} (\textbf{E}/v_0 + \textbf{B})^* \times \textbf{W}
\nonumber\\
&& + k_{eg} (\textbf{E}/v_0 + \textbf{B})^* \circ \textbf{W}_0~.
\end{eqnarray}

The force $\textbf{f} = - \textbf{n} / (2 v_0)$ in the gravitational
field and electromagnetic field can be defined from the $\textbf{n}$
.
\begin{eqnarray}
- 2 \textbf{f} = && \nabla^* w_0 + \partial_0 \textbf{w} +
(\textbf{g}/v_0 + \textbf{b})^* \times \textbf{w}/v_0
\nonumber \\
&& + \nabla^* \times \textbf{w} + w_0 (\textbf{g}/v_0 +
\textbf{b})^*/v_0
\nonumber\\
&& + k_{eg} (\textbf{E}/v_0 + \textbf{B})^* \times \textbf{W}/v_0
\nonumber\\
&& + k_{eg} (\textbf{E}/v_0 + \textbf{B})^* \circ \textbf{W}_0/v_0~,
\end{eqnarray}
where the force $\textbf{f}$ includes the inertial force, gravity,
Lorentz force, gradient of energy, and interacting force between
dipole moment with magnetic strength etc.

In the gravitational field and electromagnetic field, the above
means that Eq.(49) is much more complicated than Eq.(27), and encompasses more new force
terms about the gradient of energy etc.

In case of $\mathbb{N} = 0$ in Eq.(46), we can achieve the mass continuity equation from $n_0 = 0$ in Eq.(47), and the force equilibrium equation from $\textbf{n} = 0$ in Eq.(48) in the space $\mathbb{E}_g$ . In a similar way, we may deduce the charge continuity equation from $\textbf{N}_0 = 0$, and one new force equilibrium equation from $\textbf{N} = 0$ in the space $\mathbb{E}_e$ .

\begin{table}[t]
\caption{\label{tab:table1}The related equations when $\mathbb{F} = 0$ in the quaternion space and octonion space.}
\begin{ruledtabular}
\begin{tabular}{ll}
$ equations $                        & $ meanings $ \\
\hline
$n_0 = 0$                            & mass continuity equation \\
$\textbf{n} = 0$                     & force equilibrium equation \\
$\textbf{N}_0 = 0$                   & charge continuity equation \\
$\textbf{N} = 0$                     & force equilibrium equation \\
\end{tabular}
\end{ruledtabular}
\end{table}

\section{CONCLUSIONS}

The force can be defined from the linear momentum in the
gravitational field and electromagnetic field. The forces cover the
gravity, Lorentz force, and inertial force etc, except for the
gradient of energy. Of more important is that forces can be defined
from the energy. The latter definition of force is able to
include the gravity, Lorentz force, inertial force, gradient of
energy, interacting energy between dipole moment with electromagnetic strength, and some other new force terms, etc.

In the gravitational and electromagnetic fields, there exist
some new force terms in Eqs.(27) and (49). The term $qv_0
\textbf{B}^* \circ \emph{\textbf{I}}_0$ will cause the charged
particle to move along the magnetic force line. This is the reason
why the solar wind varies the velocity along magnetic force lines in
the sun's coronal hole as well as in the interplanetary space
between the sun and the earth.

It should be noted that the study for the new force terms examined only
some simple cases under the force definition from the linear momentum
or energy. Despite its preliminary character, this
study can clearly indicate that the gravity and Lorentz force
are only two simple force terms in the electromagnetic field and
gravitational field. Meanwhile, there exist more force terms from
the definition of energy than that from the linear momentum. For the future studies, the
research will focus on only the predictions about the velocity
variation of solar wind and charged particles in the strong electromagnetic
field.

\begin{acknowledgments}
This project was supported partially by the National Natural Science
Foundation of China under grant number 60677039.
\end{acknowledgments}

\end{document}